\documentclass[conference]{IEEEtran}
\IEEEoverridecommandlockouts
\usepackage{cite}
\usepackage{amsmath,amssymb,amsfonts}
\usepackage{algorithmic}
\usepackage{libertinus}
\usepackage{subcaption}
\usepackage{graphicx}
\usepackage{hyperref}
\usepackage{textcomp}
\usepackage{xcolor}

\usepackage{graphicx}
\usepackage{subcaption}

\def\BibTeX{{\rm B\kern-.05em{\sc i\kern-.025em b}\kern-.08em
    T\kern-.1667em\lower.7ex\hbox{E}\kern-.125emX}}
\begin{document}

\title{Quantum Repeaters Enhanced \\  by Vacuum Beam Guides\\
}
\author{
    \IEEEauthorblockN{
        Yu Gan\textsuperscript{1}, 
        Mohadeseh Azari\textsuperscript{2}, 
        Nitish Kumar Chandra\textsuperscript{2}, 
        Xin Jin\textsuperscript{1}, 
        Jinglei Cheng\textsuperscript{1}, 
        Kaushik P. Seshadreesan\textsuperscript{2}\thanks{Corresponding author: kausesh@pitt.edu},\\
        Junyu Liu\textsuperscript{1}\thanks{Corresponding author: junyuliu@pitt.edu},\\
    }
    \IEEEauthorblockA{
        \textsuperscript{1}Department of Computer Science, School of Computing \& Information, \\
        University of Pittsburgh, Pittsburgh, PA 15260, USA
    }
    \IEEEauthorblockA{
        \textsuperscript{2}Department of Informatics \& Networked Systems, School of Computing \& Information, \\
        University of Pittsburgh, Pittsburgh, PA 15260, USA
    }
}


\maketitle

\begin{abstract}
The development of large-scale quantum communication networks faces critical challenges due to photon loss and decoherence in optical fiber channels. These fundamentally limit transmission distances and demand dense networks of repeater stations. This work investigates using vacuum beam guides (VBGs)—a promising ultra-low-loss transmission platform—as an alternative to traditional fiber links.
By incorporating VBGs into repeater-based architectures, we demonstrate that the inter-repeater spacing can be substantially extended, resulting in fewer required nodes and significantly reducing hardware and operational complexity. We perform a cost-function analysis to quantify performance trade-offs across first, second, and third-generation repeaters. Our results show that first-generation repeaters reduce costs dramatically by eliminating entanglement purification. Third-generation repeaters benefit from improved link transmission success, which is crucial for quantum error correction. In contrast, second-generation repeaters exhibit a more nuanced response; although transmission loss is reduced, their performance remains primarily limited by logical gate errors rather than channel loss.
These findings highlight that while all repeater generations benefit from reduced photon loss, the magnitude of improvement depends critically on the underlying error mechanisms. Vacuum beam guides thus emerge as a powerful enabler for scalable, high-performance quantum networks, particularly in conjunction with near-term quantum hardware capabilities.

\end{abstract}


\section{Introduction}

Quantum communication marks a major advancement in information transfer, providing privacy and security by leveraging fundamental principles of quantum mechanics, such as superposition and entanglement~\cite{yang2023survey,Pirandola:20}.  It enables a range of emerging applications, including quantum key distribution~\cite{BENNETT20147,PhysRevLett.67.661}, long-distance entanglement distribution for quantum networks~\cite{Kimble2008,doi:10.1126/science.aam9288,10313694}, quantum clock synchronization~\cite{Kómár2014}, and distributed quantum sensing~\cite{Zhang_2021}. Quantum communication requires the ability to transfer quantum states between distant nodes with high fidelity. Unlike classical signals, which can be amplified or regenerated, quantum information cannot be copied due to the no-cloning theorem~\cite{wootters1982}. Consequently, quantum signals in optical fibers are vulnerable to photon loss and decoherence, limiting direct transmission distances to a few hundred kilometers. These challenges constrain the scalability of quantum networks and highlight the need for improved transmission methods and network architectures that preserve coherence over longer distances~\cite{RevModPhys.95.045006}.

To overcome the limitations posed by photon loss and decoherence over long distances, the concept of the quantum repeater was introduced~\cite{PhysRevLett.81.5932, RevModPhys.83.33}. Instead of attempting to generate entanglement directly across a long communication channel where success probabilities drop exponentially, quantum repeaters enable long-distance entanglement distribution by dividing the channel into shorter segments.  Entangled pairs are generated between neighboring nodes of these segments and extended through entanglement swapping, which combines two short-distance entangled links into one that spans a greater distance. To mitigate errors from imperfect entanglement and transmission, entanglement purification is used to distill higher-fidelity pairs from multiple noisy ones~\cite{PhysRevLett.76.722,PhysRevA.59.169}. Quantum repeaters are categorized into three generations, with the first generation employing heralded entanglement generation (HEG) and heralded entanglement purification for entanglement distribution
~\cite{muralidharan2016optimal}. Both of these operations depend on photon detection and two-way classical communication between stations to verify successful entanglement generation and coordinate subsequent operations.

While first-generation quantum repeaters offer a scalable approach to long-distance entanglement distribution, their performance is limited by several practical challenges. Among these are the probabilistic nature of entanglement generation and purification, as well as the reliance on two-way classical communication, both of which increase time and resource demands. A critical factor influencing the success rate of entanglement generation is the coupling efficiency, which describes the fraction of photons that are successfully emitted from a quantum memory, transmitted through an optical channel, and detected at the receiving end. Even modest losses in coupling efficiency can lead to a substantial drop in entanglement generation rates, compounding delays and resource requirements across the network~\cite{RevModPhys.83.33}.


 To address the challenges inherent in first-generation quantum repeaters, particularly the probabilistic nature of entanglement generation and purification as well as the dependence on long-range two-way classical communication, second- and third-generation repeater architectures have been introduced~\cite{Munro2010,Zwerger2014}. Second-generation designs leverage quantum error correction (QEC) and encoded entanglement swapping to eliminate the need for purification protocols and reduce reliance on classical communication across distant nodes. Although short-range communication is still required for heralded entanglement generation between adjacent stations, the overall entanglement distribution process becomes more efficient and exhibits reduced latency. Building on this approach, third-generation repeaters implement QEC to correct loss and operational errors, and enable high-throughput quantum communication without the need for repeated entanglement attempts or purification. Despite their conceptual advantages, both generations present significant experimental challenges. Implementing second-generation repeaters requires quantum memories with long coherence times and high-fidelity quantum gates, while third-generation systems further demand fault-tolerant quantum processors and extremely low operational error rates, which currently exceed the capabilities of available quantum technologies~\cite{muralidharan2016optimal}.

Although recent advances in quantum repeater protocols have mitigated many architectural challenges, the physical transmission medium remains a fundamental bottleneck, limiting the scalability of long-distance quantum communication. In particular, photon loss in conventional optical fibers fundamentally limits both the achievable communication range and the rate of entanglement distribution across all generations of repeater architectures. Vacuum beam guides (VBGs) present a compelling alternative by offering an ultra-low-loss channel that can maintain coherence across much greater distances~\cite{zhang2024vacuum}. Unlike optical fibers, which are affected by absorption and scattering, VBGs use a vacuum environment combined with a periodic lens system to transport photons with minimal attenuation.

The quantum network architecture proposed in Ref.~\cite{zhang2024vacuum} envisions the integration of vacuum beam guides (VBGs) with standard optical fibers and free-space quantum channels. In this architecture, VBGs enable long-distance quantum communication with significantly lower loss and reduced decoherence compared to conventional fiber links. This allows for entanglement distribution across large geographic areas at high data rates, potentially reaching the scale of terabits per second. At shorter distances, optical fiber and free-space links support communication between regional nodes and end-user devices within metropolitan or campus-scale networks. VBGs will improve the rate of entanglement generation and enhance the performance of distributed quantum protocols. Additionally, the reduction in transmission loss will help maintain error rates within the thresholds required for distributed fault-tolerant quantum computing~\cite{PhysRevResearch.5.043302}. The integration of VBGs with existing communication technologies will play a critical role in the development of scalable and high-performance quantum networks.

This study examines the application of vacuum beam guides (VBGs) as a medium for the distribution of bipartite entanglement in quantum networks. We perform a cost-function analysis encompassing all generations of quantum repeaters to address the limitations imposed by photon loss and the escalating hardware requirements in optical fiber-based quantum networks. In traditional fiber links, the probability of successfully establishing a quantum link diminishes exponentially with distance, necessitating the frequent installation of repeaters and exacerbating system complexity. 
Our analysis suggests that VBG networks incur lower costs while delivering equivalent performance and surpassing the capabilities of optical fiber in fixed architectures. By enhancing transmission fidelity, VBGs significantly contribute to the facilitation of efficient, scalable quantum communication compatible with forthcoming quantum technologies.

The remainder of this paper is organized as follows. Section~\ref{bg} provides the theoretical background relevant to quantum repeater architectures. In Section~\ref{sec:cost_function}, we describe the cost function used to compare different quantum repeater architectures. Section~\ref{results} presents the main results of our study, including the cost-function analysis and performance evaluation across various repeater generations. Finally, Section~\ref{conclusion} summarizes the key findings and outlines future directions for integrating VBGs into large-scale quantum communication networks.

\section{Background Theory}\label{bg}

\subsection{Heralded Entanglement Generation}



Heralded entanglement generation (HEG) is a technique for establishing entanglement between distant nodes in quantum repeater-based networks. In a typical implementation, each node prepares a local entangled state between a stationary qubit, such as an atomic spin, and an emitted photonic qubit. The photons are then transmitted through optical fibers to a central measurement station, where they interfere at a beam splitter. A joint measurement, usually a Bell-state measurement, is performed on the photons, and specific detection outcomes indicate the successful entanglement of the remote stationary qubits~\cite{PhysRevLett.81.5932,Lvovsky2009}. This method does not require direct interaction between the quantum memories, and is compatible with various platforms including trapped ions, neutral atoms, and solid-state systems. However, due to losses in the optical channel and the non-unit efficiency of single-photon detectors, the success probability is low. Consequently, multiple attempts are typically required, and quantum memories must store qubits until a successful heralding event occurs. This leads to increased latency in the entanglement distribution process~\cite{Duan2001}.

The efficiency of heralded entanglement generation (HEG) depends on photon collection efficiency, transmission losses, and detector performance. Losses arising from fiber attenuation, mode mismatch, and limited detector efficiency reduce the probability of successfully establishing entanglement. Therefore, improving coupling efficiency and minimizing transmission losses are essential for enhancing entanglement generation rates in practical quantum networks~\cite{Maunz2007, RevModPhys.83.33}.


\subsection{Heralded Entanglement Purification}

\begin{figure}[hbt!]
    \centering
    \includegraphics[width=0.98\columnwidth]{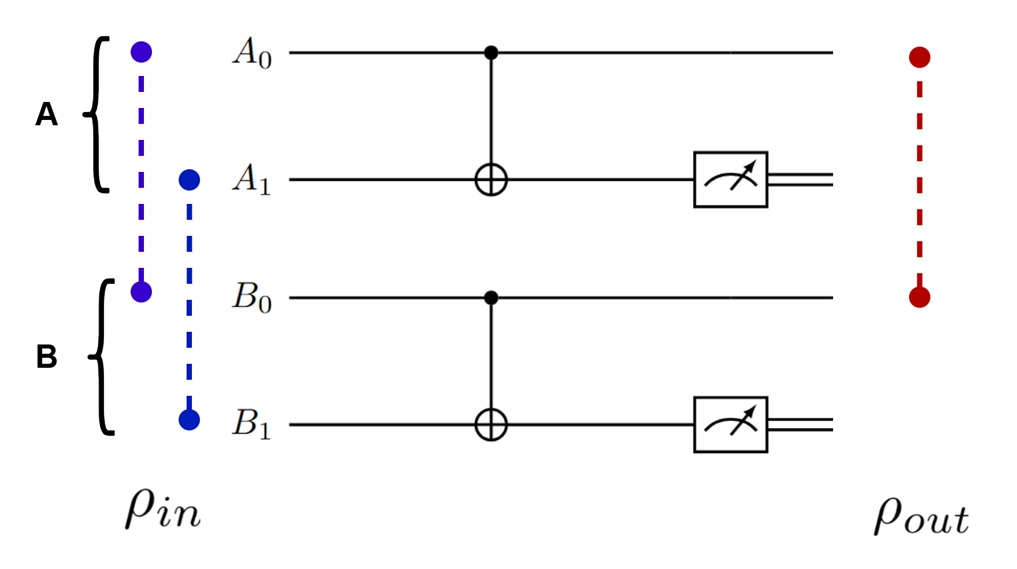}
    \caption{Schematic of an entanglement distillation protocol. We consider two remote parties, $A$ and $B$, engaged in an entanglement distillation protocol. They share two two-qubit states, \( \rho_{A_0B_0} \) and \( \rho_{A_1B_1} \), both initialized in a noisy Bell state. User A holds an entangled pair \( A_0A_1 \), and user B holds a corresponding pair \( B_0B_1 \), together representing the noisy entangled input state \( \rho_{\text{in}} \). Both parties perform local operations: a CNOT gate is applied between their respective qubits \( A_0 \rightarrow A_1 \) and \( B_0 \rightarrow B_1 \), followed by a measurement on the target qubits \( A_1 \) and \( B_1 \). Postselection is performed based on the measurement outcomes, and if successful, the remaining pair \( A_0B_0 \) forms a new entangled state \( \rho_{\text{out}} \) with improved fidelity. This process can be repeated iteratively to approach near-maximally entangled states.}

    \label{distillation}

\end{figure}

Entanglement purification is used for enhancing the fidelity of entangled states distributed over noisy quantum channels. It enables the extraction of high-fidelity entangled pairs from multiple imperfect ones~\cite{PhysRevLett.76.722,PhysRevA.59.169}. Among the earliest entanglement purification protocols is the BBPSSW protocol~\cite{PhysRevLett.76.722}, which is used for Werner states that are mixtures of Bell states affected by isotropic noise. The procedure involves applying CNOT gates to two shared entangled pairs, followed by measurements and classical communication~(See~Fig.\ref{distillation}). If both parties obtain matching measurement outcomes, we retain the control pair, which now has higher fidelity. This approach improves entanglement quality through postselection but may be limited when dealing with more general noise that deviates from the Werner form. To overcome these limitations, the DEJMPS protocol~\cite{PhysRevLett.77.2818} was proposed as an alternative for Bell-diagonal states. It builds upon the BBPSSW framework but adds an initial step involving local basis rotations to symmetrize the state. These transformations redistribute weight from less favorable components to the target Bell state, increasing the likelihood of successful purification. This adjustment enhances the protocol’s efficiency, particularly in scenarios where the noise is non-uniform, resulting in faster convergence and higher fidelity after fewer rounds.

The BBPSSW and DEJMPS protocols are fundamental to entanglement purification in quantum repeater architectures. These protocols ensure that only high-fidelity entangled pairs are used in subsequent operations such as entanglement swapping. Entanglement purification can also be employed in satellite-based quantum communication and complex quantum network architectures~\cite{doi:10.1126/science.aan3211,PhysRevResearch.5.033171}. Its practical appeal lies in its compatibility with existing optical technologies and its relatively modest resource requirements compared to full-scale quantum error correction. However, HEP alone cannot eliminate all noise sources, it is often combined with additional techniques such as entanglement pumping to improve scalability and performance~\cite{Dür_2007}.


\subsection{Quantum Error Correction (QEC)}




Quantum error correction (QEC) is essential for achieving reliable quantum communication in second and third generation quantum repeater architectures. QEC introduces redundancy by encoding a single logical qubit into a block of multiple physical qubits, enabling the system to detect and correct arbitrary single-qubit errors. Among the most widely used classes of QEC codes are the Calderbank-Shor-Steane (CSS) codes~\cite{gottesman2002introduction,steane1999enlargement}, which allow repeater nodes to correct errors that occur during qubit transmission, quantum gate operations, and memory storage. Fault-tolerant operations can be achieved using transversal gates, which apply logical operations independently on corresponding physical qubits across code blocks. Additionally, syndrome extraction circuits can identify the presence and type of errors without measuring or collapsing the encoded quantum information~\cite{knill1998resilient,lidar2013quantum}.

While quantum error correction (QEC) is theoretically capable of enabling reliable long-distance quantum communication, its practical viability depends critically on the physical characteristics of the communication channel. Integrating QEC with low-loss quantum channels can significantly reduce the need to encode quantum information into large blocks of physical qubits, thereby easing resource requirements. The effectiveness of QEC is governed by two key thresholds. The first is the fault-tolerance threshold, which sets an upper bound on gate and memory error rates; below this threshold, logical errors can be exponentially suppressed by increasing the code size~\cite{Dür_2007}. The second is the loss-tolerance threshold, which limits the maximum number of photon losses that a code can withstand before successful decoding becomes impossible~\cite{PhysRevLett.105.200502}. These thresholds are particularly relevant in quantum communication networks, where transmission loss is often the dominant source of error.

\subsection{Three generations of Quantum Repeaters}

Here, we briefly outline the three generations of quantum repeaters, highlighting their underlying principles and the advancements they provide in addressing the challenges of long-distance quantum communication.
\vspace{3pt}

\begin{itemize}
    \item  \textit{First-generation quantum repeaters : } These generation quantum repeaters address photon loss by dividing the total communication distance into shorter segments, within which entangled states are generated between neighboring nodes using \textit{heralded entanglement generation}. In this process, photons emitted from both nodes interfere at a central station, and a coincident detection event heralds the successful creation of an entangled pair. These short-range entangled pairs are extended across longer distances via \textit{entanglement swapping}, where a Bell-state measurement performed at an intermediate node projects the outermost nodes into an entangled state. At each nesting level $k$, two links of length $L_{k-1}$ are combined to form a longer link of length $L_k = 2^k L_0$, where $L_0$ is the elementary segment length. This process is repeated recursively until entanglement spans the entire distance. To preserve entanglement fidelity during this process, \textit{entanglement purification} is employed. Purification typically involves local quantum operations and two-way classical communication to identify which pairs can be retained. While first-generation repeaters overcome the exponential scaling of direct transmission, they still incur substantial overhead due to the probabilistic nature of entanglement generation and purification, as well as the reliance on classical communication at each step, which increases latency and resource demands.

\vspace{3pt}
\item{\textit{Second Generation Quantum Repeaters :}} Second-generation quantum repeaters improve the scalability of long-distance quantum communication by addressing both photon loss and operational errors. While first-generation repeaters rely on heralded entanglement generation (HEG) and purification to mitigate photon loss, they do not correct errors introduced by imperfect gates and measurements. To overcome this limitation, second-generation architectures integrate quantum error correction (QEC) protocols, which allow for the detection and correction of errors that accumulate during multi-step quantum operations. Logical qubit states, such as \( |0\rangle_L \) and \( |+\rangle_L \), are encoded using quantum error-correcting codes, typically from the CSS family. To establish entanglement between neighboring stations, teleportation-based non-local CNOT operations are performed on each physical qubit in the encoded block, using entangled pairs generated through HEG. A major advantage of this approach is that it removes the need for probabilistic entanglement purification, which in first-generation repeaters required two-way classical communication to compare measurement outcomes and discard unsuccessful attempts. This significantly reduces latency and increases the rate of entanglement distribution. As long as the total noise remains below the threshold of the QEC code, high-fidelity entanglement can be reliably maintained over long distances without the need for repeated purification. This makes second-generation repeaters a strong candidate for building practical and scalable quantum communication networks.
\vspace{3pt}
\item{\textit{Third-generation quantum repeaters :} These architectures adopt a fundamentally distinct strategy from earlier designs by eliminating the need for intermediate entanglement distribution or purification steps. Instead of generating entangled pairs between nodes, quantum information is transmitted directly through the network in a form protected by quantum error correction (QEC). In this model, the quantum state is encoded across a block of physical qubits capable of withstanding both photon loss and operational errors. As long as the noise encountered during transmission remains within the correctable threshold of the code, the destination node can accurately recover the original state from the received qubits. One of the most significant advantages of this architecture is its reliance on one-way classical communication, removing the delays associated with the two-way signaling required in previous repeater generations. Each repeater node independently performs decoding and re-encoding operations without the need for feedback from neighboring stations. This approach significantly improves communication rates, as the speed is governed primarily by local processing rather than classical communication delays. As a result, third generation repeaters offer a promising pathway toward scalable, high-rate quantum communication networks, although their realization remains challenging due to the limitations of current experimental capabilities.}

\end{itemize}

\begin{figure*}[t]
    \centering
    \includegraphics[width=0.90\textwidth]{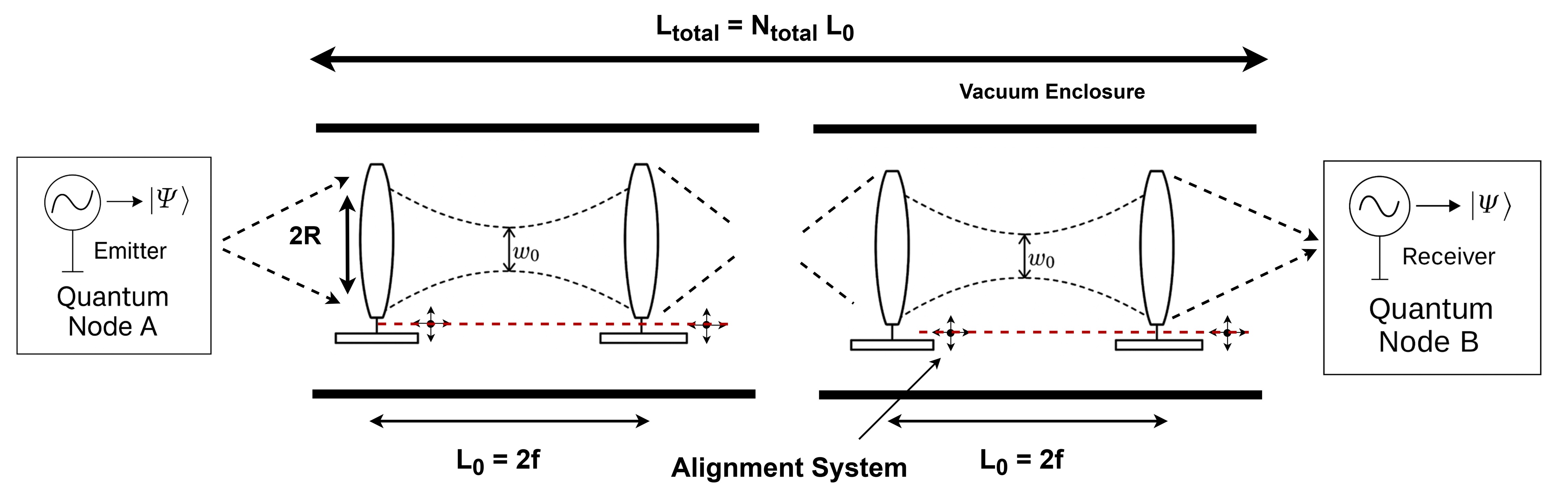} 
    \caption{Schematic illustration of a vacuum beam guide (VBG)-based quantum communication system, where Quantum Node~A is configured as an emitter and Quantum Node~B as a receiver. The quantum state \(|\Psi\rangle\), transmitted from Quantum Node~A, propagates through a sequence of lenses with identical focal length \(f\) and radius \(R\), spaced periodically within a vacuum enclosure. This configuration forms the VBG, which supports low-loss transmission of a fundamental Gaussian mode with waist \(w_0\) over a total distance \(L_{\text{total}} = N_{\text{total}} L_0\), where \(L_0 = 2f\). An active alignment system ensures mode stability across all \(N_{\text{total}}\) sections. At Node~B, the incoming photon is collected by receiver and directed into a quantum memory for further quantum processing. 
     }
    \label{VBG}
\end{figure*}


\subsection{Vacuum Beam Guides}
Efficient entanglement distribution over long distances is a fundamental challenge in developing quantum networks. Optical fibers, the traditional medium for transmitting quantum states, suffer from intrinsic photon loss that grows exponentially with distance. This leads to strict limitations on achievable entanglement distribution rates and necessitates the deployment of a large number of intermediate quantum repeaters.
The theoretical upper limit on the rate of entanglement distribution over a lossy bosonic channel without repeaters is set by the Pirandola-Laurenza-Ottaviani-Banchi (PLOB) bound~\cite{pirandola2017fundamental}. For an optical fiber channel with transmissivity $\eta$, the PLOB bound restricts the secret-key capacity per optical mode to $-\log_2(1-\eta)$ bits. Since fiber transmissivity decreases exponentially with distance, this imposes severe constraints on repeaterless quantum communication, and even with repeaters, the total resource costs become substantial.

Recent developments in free-space optics propose Vacuum Beam Guides (VBGs) as an alternative transmission medium to mitigate these limitations~\cite{zhang2024vacuum}. In a VBG, photons propagate through evacuated tubes equipped with periodic lens arrays to counteract diffraction and maintain beam quality over extremely long distances (See~Fig.\ref{VBG}). This approach results in attenuation lengths on tens of thousands of kilometers, dramatically surpassing the typical $L_{\mathrm{att}} \sim 20$~km of standard optical fibers.
The advantages of employing VBGs are significant. First, the extended attenuation length enables the placement of repeater stations much farther apart, drastically reducing the number of repeaters required to span a given total distance. This directly impacts the cost function, defined as the resource overhead (qubits $\times$ time) per secret bit per kilometer. By reducing the repeater count and associated operational overhead, VBGs offer the potential for lower cost coefficients compared to fiber-based networks.
Additionally, VBGs provide a ground-based solution less sensitive to atmospheric variations than satellite-based free-space communication. Their high transmission capacities enable entanglement distribution rates that could rival or surpass those achievable by conventional technologies.

Despite their advantages, the use of vacuum beam guides (VBGs) introduces critical trade-offs.
Their performance  is susceptible to the precise alignment of optical elements. Over long distances, environmental disturbances such as seismic activity, thermal expansion, and mechanical drift can degrade alignment and reduce system efficiency. Furthermore, constructing and maintaining evacuated tubes over continental scales entail substantial infrastructural and economic investments. These factors introduce operational challenges and ongoing maintenance costs that must be carefully weighed against the benefits.
In summary, VBGs represent a promising technological advancement for large-scale quantum networks, offering the possibility of entanglement distribution over unprecedented distances with reduced repeater requirements. Nevertheless, a careful cost-benefit analysis must be conducted, considering the dramatic reduction in loss and the engineering complexity, alignment stability, and infrastructural demands associated with implementing VBGs at scale. In this work, we present a cost-function-based analysis comparing fiber-based and vacuum-based quantum repeater architectures across all three generations.

\begin{figure*}[t]
    \centering
    \includegraphics[width=.8\linewidth]{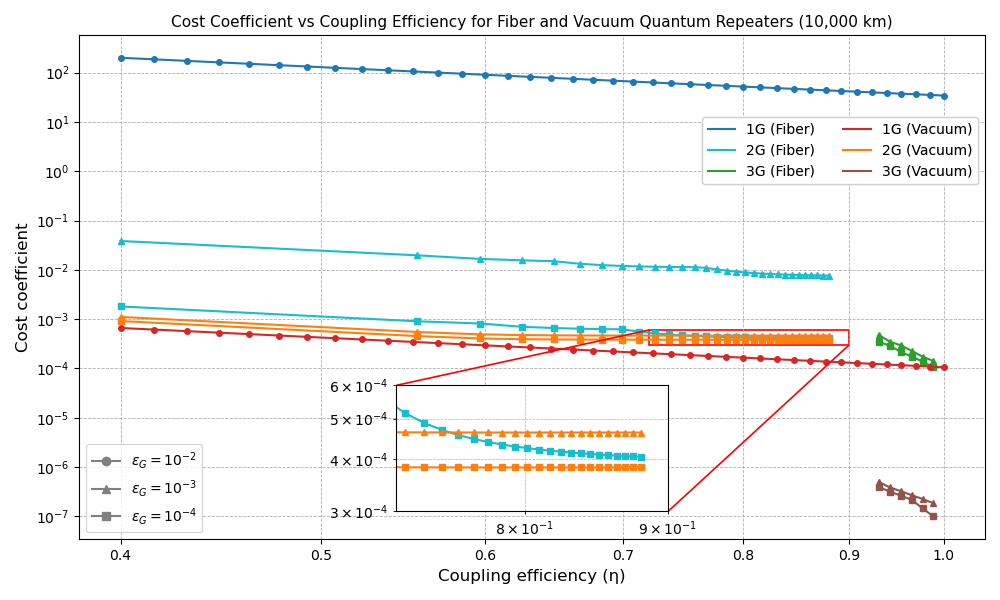}
    \caption{Cost coefficient versus coupling efficiency ($\eta$) for three generations of quantum repeaters (QRs), evaluated over a total distance of 10{,}000~km. Both fiber-based and vacuum-based repeater architectures are compared for each generation. Marker shapes indicate the logical gate error rate: circles for $\varepsilon_G = 10^{-2}$ (1G only), triangles for $\varepsilon_G = 10^{-3}$, and squares for $\varepsilon_G = 10^{-4}$. Fiber-based systems assume typical optical attenuation (e.g., 20~km), while vacuum systems use significantly higher attenuation lengths (e.g., 42{,}000~km) to reflect lower channel loss.}
    \label{fig:overall_reults}
\end{figure*}

\begin{figure*}[t]
    \centering

    \begin{minipage}{0.65\linewidth}
        \centering
        \includegraphics[width=\linewidth]{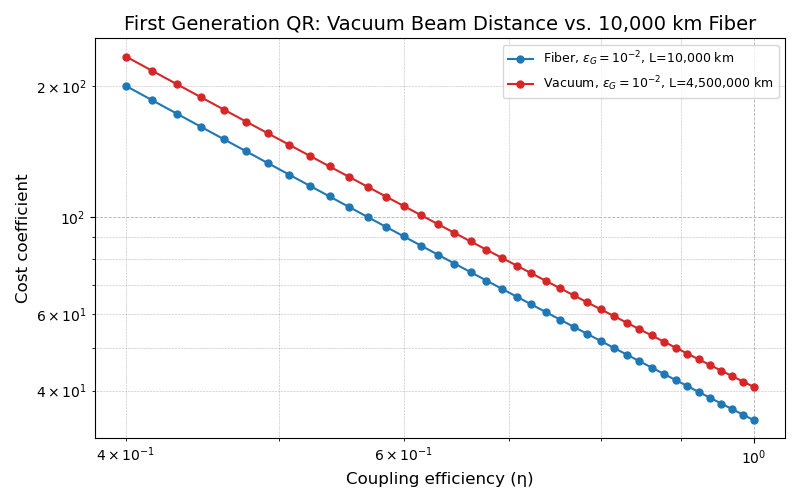}
    \end{minipage}
    \captionsetup{width=\textwidth}
    \caption{Cost coefficient versus coupling efficiency ($\eta$) for first-generation quantum repeaters using fiber and vacuum beam transmission. Both setups use a gate error rate of $\varepsilon_G = 10^{-2}$, with fiber operating over 10{,}000~m and vacuum over 4{,}500{,}000~m (450$\times$ longer).}
    \label{fig:first_gen_approx}

    \vspace{2em}

    \begin{minipage}{0.65\linewidth}
        \centering
        \includegraphics[width=\linewidth]{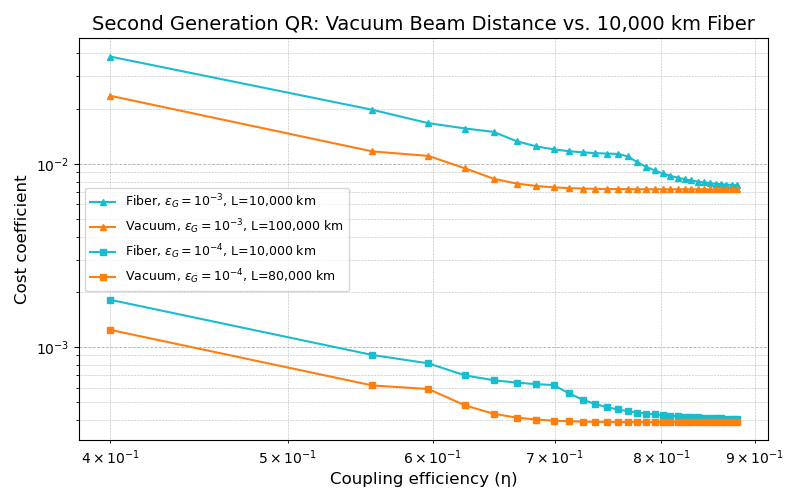}
    \end{minipage}
    \captionsetup{width=\textwidth}
    \caption{Cost coefficient versus coupling efficiency ($\eta$) for second-generation quantum repeaters using fiber and vacuum beam guide. Fiber-based QRs are constrained to 10,000~km, while VBG systems support distances up to 100,000~km with similar or lower cost.}
    \label{fig:second_gen_approx}
\end{figure*}

\section{Cost Function}\label{cost_function}
\label{sec:cost_function}

Designing a cost function for quantum repeaters that distribute Bell pairs requires capturing both hardware resource demands and the quality of entanglement distribution. The cost should integrate contributions from the number of quantum memory qubits, gate operation inefficiencies, and classical communication overhead. 
The repeater's performance can be evaluated in terms of the secure key generation rate, which depends on the probability of successfully generating entanglement across all elementary links, the distance over which the entangled state is distributed, and the rate at which secure bits are produced per successful entanglement generation. Specifically, the secure rate decreases exponentially with the number of elementary links, as each link independently contributes to the probability of overall success. 

In addition, the total time required for entanglement distribution includes the intrinsic time per elementary link and the classical communication delay. Consequently, the achievable rate reflects a trade-off between success probability, link distance, and timing overhead. Performance metrics such as the fidelity of the distributed entangled states, the total latency, and the overall success probability must be considered as functions of physical parameters, including measurement inefficiencies, gate errors, and memory coherence times.
We define the unified cost function as
\begin{equation}
    C = R_{\text{dep}} \times \frac{\textrm{number of qubits per repeater}}{\textrm{repeater spacing (km)}},
\end{equation}
where $R_{\text{dep}}$ is the bipartite entanglement achievable rate.

This form ensures that the cost increases with greater resource consumption or longer generation time, and decreases with higher fidelity and success probability. High-fidelity or high-rate architectures are naturally favored. The fidelity--rate trade-off is embedded in the cost function: improving fidelity often reduces the rate and vice versa, allowing different applications to adjust the relative weighting based on their operational requirements.

Importantly, this cost function remains independent of the specific repeater generation or protocol. It can compare first-generation heralded entanglement swapping, second-generation purification-assisted repeaters, and third-generation error-corrected schemes on equal footing. It also generalizes to multipartite entanglement distribution, where $N$ denoting the number of qubit in the entangled state we intent to share appears implicitly through resource scaling and performance degradation.

Our framework unifies and extends previous cost metrics found in the literature, including qubit-time per secure bit and node-weighted cost per key rate. By combining hardware and performance into a single figure of merit, the proposed cost function enables systematic optimization and comparison of quantum repeater architectures for scalable quantum networks.

\section{Results}\label{results}


Long-distance quantum communication is fundamentally constrained by photon loss in transmission media. Although optical fibers are widely used, their high attenuation, approximately \( L_{\mathrm{att}} \sim 20\,\mathrm{km} \) at telecom wavelengths, necessitates closely spaced repeater stations. Vacuum Beam Guides (VBGs) offer a promising alternative by enabling photon transmission with minimal scattering and absorption. With an effective attenuation length reaching up to \( L_{\mathrm{att}} \sim 42{,}000\,\mathrm{km} \), VBGs allow for significantly greater spacing between repeater nodes. 


We begin by evaluating the performance of VBGs in first-generation quantum repeater architectures and extend the comparison to second and third-generation schemes. We analyze how the optimized cost coefficient \( C' \) varies with coupling efficiency \( \eta \), under fixed gate error rates \( \varepsilon_G \) and a total communication distance of \( L_{\mathrm{tot}} = 10{,}000\,\mathrm{km} \). Figure~\ref{fig:overall_reults} presents \( C' \) as a function of \( \eta \) for QRs implemented using either optical fiber or VBGs. Each generation is shown under relevant gate error rates: \( \varepsilon_G = 10^{-2} \) for 1G, and \( \varepsilon_G = 10^{-3} \) and \( 10^{-4} \) for 2G and 3G.
In most cases, VBG based implementations yield significantly lower cost coefficients than their fiber-based counterparts, even as \(\eta\) increases. This indicates that VBGs can substantially reduce the resource overhead required for entanglement distribution at long distances. Interestingly, the cost coefficient for the first-generation VBG configuration is observed to be lower than that of second-generation VBGs in certain regimes. This stems from the optimal configuration identified for the first-generation case, which employs a nesting level of \( N_{nest} = 1 \) and no entanglement purification (i.e., \( \overrightarrow{M} = [0, 0] \)), thereby eliminating additional purification overhead. In contrast, second-generation repeaters inherently require purification and quantum error correction to counteract gate noise, especially in fiber-based implementations. As shown by the difference between \( \varepsilon_G = 10^{-3} \) and \( 10^{-4} \), the performance gap between VBG and fiber-based architectures widens with increasing noise, reinforcing the utility of VBGs in low-fidelity settings. Nonetheless, VBGs do not universally outperform fiber. For instance, under high coupling efficiency and low gate error (e.g., \( \varepsilon_G = 10^{-4} \)), second-generation fiber-based QRs may achieve slightly lower cost coefficients than their VBG counterparts. This underscores that the benefits of VBGs are most prominent when gate noise is non-negligible and/or \(\eta\) is moderate. From an architectural standpoint, first- and second-generation QRs are most effective in mid-range coupling regimes. In contrast, third-generation QRs exhibit a distinct behavior: their performance is highly sensitive to \(\eta\), becoming cost-effective only when coupling efficiency is high. This aligns with the third generation’s reliance on local gate operations and fault-tolerant error correction, which demand high transmission efficiency to amortize the associated overhead.

In Fig.~\ref{fig:first_gen_approx} and Fig.~\ref{fig:second_gen_approx}, we further examine scenarios in which vacuum beam guides (VBGs) are employed over significantly extended distances—specifically, $L_{\mathrm{tot}} = 4{,}500{,}000\,\mathrm{km}$ for first-generation and $L_{\mathrm{tot}} = 100{,}000\,\mathrm{km}$ and $80{,}000\,\mathrm{km}$ for second-generation repeaters. Remarkably, the resulting cost coefficients remain comparable to those achieved by optical fiber repeaters operating over just $10{,}000\,\mathrm{km}$. This observation highlights a key advantage of VBG-based architectures: they enable quantum repeater networks to maintain low resource costs even over interplanetary-scale distances, substantially extending the feasible communication range without requiring additional repeater complexity.
Overall, our results demonstrate that the superior attenuation properties of VBGs enable significantly larger inter-repeater spacings while maintaining low cost coefficients, thus offering a promising path toward scalable long-distance quantum communication.
We further evaluated the performance of Vacuum Beam Guides (VBGs) by comparing their impact on second and third-generation quantum repeaters. The comparison used the resource overhead metric, defined as the product of the number of qubits and the communication time required to create one secret bit over a 1 km distance, following the cost analysis framework in~\cite{muralidharan2016optimal}.
In the third generation of quantum repeaters, where end-to-end entanglement is distributed by directly transmitting encoded qubits, the use of VBGs led to a substantial improvement in resource overhead. This improvement is primarily attributed to the significant enhancement in the success probability \( p \) of detecting X/Z errors in the transmitted resource qubits. As the attenuation length \( L_{\mathrm{att}} \) in VBGs is four orders of magnitude larger than that of optical fibers, the probability of successfully generating entanglement between two adjacent memory qubits in a single trial is correspondingly improved. This probability is given by
\[
p = \frac{1}{2} \eta_c \exp\left(-\frac{L_0}{L_{\mathrm{att}}}\right),
\]
where \( \eta_c \) is the coupling efficiency and \( L_0 \) is the inter-repeater spacing. The higher value of \( p \) enhances the success probability \( P_{\mathrm{success}} \) of establishing entanglement across adjacent repeaters after multiple rounds of entanglement generation:
\[
P_{\mathrm{success}} = 1 - (1 - p)^{M \times n_{\mathrm{EG}}},
\]
where \( M \) is the number of memory qubits and \( n_{\mathrm{EG}} \) is the number of elementary generation attempts.
In contrast, for second-generation repeaters, which rely on heralded entanglement generation between unencoded qubits followed by encoded teleportation, the transition from optical fiber to VBGs resulted in only modest performance gains. Although the attenuation length increases dramatically, the dependence of \( P_{\mathrm{success}} \) on the parameter \( p \) is such that the overall improvement in success probability is minimal. As a result, the cost function optimization shows only a marginal reduction in resource overhead for second-generation repeaters when using VBGs compared to fibers.
Overall, the results confirm that VBGs significantly enhance the performance of third-generation quantum repeaters by allowing longer inter-repeater spacings and higher success probabilities, thus reducing the required number of repeaters and the total resource overhead. However, the benefits for second-generation repeaters are relatively limited, suggesting that VBGs are most impactful in architectures that leverage direct transmission of encoded quantum information.

\subsection{Optimization Analysis}
\label{sec:optimization}

The optimization analysis, summarized in Fig.~\ref{fig:overall_reults}, reveals the distinct impacts of the transmission medium on different generations of quantum repeaters. Several key trends emerge when substituting optical fiber with vacuum beam guides, which offer attenuation lengths on the order of $42,000$~km.

First, the bottleneck is the link transmission success probability for third-generation quantum repeaters, which rely on quantum error correction (QEC) to address both loss and operational errors. As shown in Fig.~\ref{fig:overall_reults}, vacuum beam guides significantly improve this probability, substantially reducing the cost coefficient. Consequently, third-generation architectures benefit the most from a high-transmission environment, especially when coupling efficiencies are high.
Second, in the case of second-generation repeaters, although vacuum beam guides improve transmission loss, the dominant bottleneck becomes the gate error rate rather than the channel loss. Because second-generation protocols still use heralded entanglement generation (HEG) for loss errors, improvements in the medium do not dramatically lower the cost unless gate operations are highly reliable. This is evident in the zoomed inset of Fig.~\ref{fig:overall_reults}, where fiber-based architectures with lower gate errors outperform vacuum-based architectures with higher gate errors.
Third, the vacuum beam guide offers a particularly striking advantage for first-generation quantum repeaters. The increased transmission success leads to an optimal nesting level of zero, effectively eliminating the need for entanglement purification. As a result, direct entanglement distribution becomes feasible, drastically improving the achievable rate and substantially lowering the cost coefficient, as illustrated by the sharp drop in Fig.~\ref{fig:overall_reults}.

Overall, these findings emphasize that the benefits of improved transmission mediums, such as vacuum beam guides, are most pronounced in third-generation repeaters, where one-way communication and QEC are leveraged. First-generation architectures also benefit dramatically by bypassing purification, while second-generation repeaters exhibit a more nuanced response where gate performance remains the critical limiting factor. These results guide designing future large-scale quantum networks and integrating vacuum-based transmission technologies.

\section{Discussion and Conclusion}\label{conclusion}
Our results demonstrate that integrating Vacuum Beam Guides into quantum repeater architectures offers substantial advantages in terms of cost efficiency and scalability, particularly for long-distance quantum communication. The improvement is most dramatic in third-generation repeaters, where encoded qubits are directly transmitted between nodes. In contrast, first and second-generation repeaters show only moderate or limited improvement, respectively, due to their reliance on heralded entanglement generation and purification, which do not benefit exponentially from increased attenuation length. Notably, the first-generation VBG architecture can even outperform second-generation implementations in specific settings where purification is avoided, due to its simplified structure and minimal resource requirements.

Another important insight is the sensitivity of each generation to coupling efficiency. Second-generation QRs remain effective in moderate \(\eta\) regimes, but their advantages diminish at high coupling efficiency when gate error rates are low. On the other hand, third-generation QRs are only cost-effective in high-\(\eta\) regimes, reflecting their dependence on fault-tolerant error correction and local operations. These trends suggest that future quantum network designs must carefully consider the target physical parameters when selecting QR architecture and transmission medium.


\section*{Acknowledgment}
We thank Chaohan Cui, Saikat Guha, Yuexun Huang, Liang Jiang, Prashant Krishnamurthy and Pei Zeng for helpful discussions. JL is supported in part by the University of Pittsburgh, School of Computing and Information, Department of Computer Science, Pitt Cyber, and by NASA under award number 80NSSC25M7057. JL and KPS are supported by the Pittsburgh Quantum Institute Community Collaboration Award. This research used resources of the Oak Ridge Leadership Computing Facility, which is a DOE Office of Science User Facility supported under Contract DE-AC05-00OR22725.

\bibliographystyle{IEEEtran}
\bibliography{references}
\end{document}